\documentclass[prb,twocolumn,showpacs,preprintnumbers,amsmath,amssymb]{revtex4}

\usepackage{graphicx}
\usepackage{dcolumn}
\usepackage{bm}

\begin{document}


\title{Enhanced superconducting pairing interaction in indium-doped tin telluride}

\author{A. S. Erickson,$^1$ J. -H. Chu,$^1$ M. F. Toney,$^2$ T. H. Geballe,$^1$ I. R. Fisher$^1$} 
\affiliation{$^1$Department of Applied Physics and Geballe Laboratory for Advanced Materials, Stanford University, CA 94305.\\
$^2$Stanford Synchrotron Radiation Laboratory, Menlo Park, CA 94025.
}

\date{\today}

\begin{abstract}
The ferroelectric degenerate semiconductor Sn$_{1-\delta}$Te exhibits superconductivity with critical temperatures, $T_c$, of up to 0.3 K for hole densities of order 10$^{21}$ cm$^{-3}$.  When doped on the tin site with greater than $x_c$ $= 1.7(3)\%$ indium atoms, however, superconductivity is observed up to 2 K, though the carrier density does not change significantly.  We present specific heat data showing that a stronger pairing interaction is present for $x > x_c$ than for $x < x_c$.  By examining the effect of In dopant atoms on both $T_c$ and the temperature of the ferroelectric structural phase transition, $T_{SPT}$, we show that phonon modes related to this transition are not responsible for this $T_c$ enhancement, and discuss a plausible candidate based on the unique properties of the indium impurities.

\end{abstract}

\pacs{74.20.Mn, 74.62.Dh, 74.70.Ad}
\maketitle
\section{Introduction}

When doped with at least $10^{20} cm^{-3}$ vacancies on the Sn site, Sn$_{1-\delta}$Te superconducts, with critical temperatures remaining below 0.3 K for carrier densities up to $2 \times 10^{21}cm^{-3}$, agreeing with BCS predictions.\cite{allen1969}  When doped on the Sn site with In, however, superconductivity is found as high as 2 K, though the carrier concentration remains similar.\cite{bushmarina1986} Significantly, In is known to skip the $+2$ valence, in favor of $+1$ and $+3$, in most materials (e.g. InCl and InCl$_3$).  Similar systems of degenerate semiconductors doped with valence-skipping elements (e.g. Pb$_{1-x}$Tl$_x$Te\cite{yana} and BaPb$_{1-x}$Bi$_x$O$_3$\cite{sleight}) are also known to superconduct at anomalously high temperatures, given their low carrier concentration.  In the case of Pb$_{1-x}$Tl$_x$Te, the Tl dopant atom has been found to play a critical role in enhancing $T_c$, intimately related to its unique mixed valent state.\cite{yana}  This work seeks to investigate the role of In dopant atoms in enhancing $T_c$ in Sn$_{1-\delta-x}$In$_x$Te.

In addition to superconductivity, tin telluride also exibits a ferroelectric structural phase transition (SPT).  The transition temperature, $T_{SPT}$, decreases with increasing hole concentration, $p$, from $\sim$ 100 K at $p \sim 10^{20} cm^{-3}$ to 0 K at $p \sim 10^{21} cm^{-3}$, due to Thomas-Fermi screening of the ferroelectric phonon modes by the electronic carriers.\cite{kobayashi1976} Near $T_{SPT}$, resistive scattering is observed to increase,\cite{kobayashi1976} which has been associated with strong electron-phonon coupling to the TO phonon mode responsible for the ferroelectric transition.\cite{katayama1980} Above $T_{SPT}$, the material has a cubic rocksalt structure, while the low temperature structural phase has been characterized as rhombohedral by powder x-ray diffraction.\cite{muldwater1973} The Bragg reflections of this structure differ from the high temperature cubic phase by a splitting of (hk0) reflections into two peaks of equal intensity, while (h00) peaks remain unsplit. The magnitude of this peak splitting provides a natural order parameter for the phase transition.


In the present work, we study the effect of In dopant concentration on the structural phase transition in Sn$_{1-\delta-x}$In$_x$Te by tracking the resistive signal in single crystals, and linshapes of (hk0) reflections in high resolution powder x-ray diffraction as a function of temperature.  We find $T_{SPT}$ to be suppressed smoothly with increasing indium content in a manner consistent with the increase in carrier concentration. However, we find a threshold rise in $T_c$ from $T_c$ $<$ 0.35 K for $x$ $<$ $x_c$ $=$ 1.7(3)\% to $T_c$ $>$ 0.8 K for $x$ $>$ $x_c$ that is not correlated with a similar anomaly in $T_{SPT}$. This suggests that the soft phonon mode associated with the ferroelectric transition is not responsible for enhanced superconductivity in In-doped Sn$_{1- \delta}$Te.  Furthermore, comparing $T_c$ to the density of states at the Fermi level, $N(0)$, extracted from specific heat measurements reveals enhanced superconducting pairing strength for $x$ $>$ $x_c$ and a further variation of pairing strength with $x$ as $x$ is increased beyond $x_c$.  Possible candidates for the source of this enhanced pairing are discussed.

\section{Experimental Methods}
Single crystals of Sn$_{1-\delta-x}$In$_x$Te were grown from a Sn-rich melt.  This growth process is preferable to many other methods for these experiments because it leads to relatively low amounts of Sn vacancies of order $10^{20}$ cm$^{-3}$, which leaves $T_{SPT}$ high enough in temperature to observe its dependence over a wide range of $x$.  Elemental starting materials in the ratio $85\times(1-x)$:$85\times x$:15 ($x =$ 0 to 0.18) Sn:In:Te were placed in an alumina crucible, sealed under vacuum in a quartz ampoule, and heated to 725 C.  The melt was cooled at a rate of 2 degrees per hour to 400 C, and the Sn-rich flux was decanted.  The resulting single crystals were annealed at 550 C for 48 hours to improve crystal quality, measured by sharpness of the SPT transition in resistivity data.  The composition was measured by electron microprobe analysis (EMPA), using elemental Sn, In, and Tellurium, and PbTe standards. Resulting In concentrations of crystals grown in this method are approximately 60\% of the initial melt concentration.  Material of measured In concentration 0\%, 0.47(4)\%, 1.0(1)\%, 1.4(1)\%, 2.1(1)\%, 2.7(1)\%, 3.4(1)\%, 4.4(1)\%, 5.1(1)\%, and 9.9(1)\% were used in this study. Data are also included for one sample not characterized by EMPA, with an estimated In concentration of 2.3\% 
, based on initial growth conditions.

Bars were cleaved from single crystals for electrical transport measurements.  Resistivity data were obtained at frequencies of 13.7 Hz or 37 Hz and current densities of order 100 mA/cm$^2$, along the [100] direction.  Data taken below 1.8 K were also taken at lower current densities to check for heating.
Hall data were collected using a Quantum Design Physical Properties Measurement System, at a frequency selected to reduce noise (either 37 Hz, 47 Hz, or 53 Hz) and typical current densities of order 100 mA/cm$^2$.  The Hall coefficient, $R_H$, at a temperature of 5 K was obtained from linear fits to the transverse voltage in fields from -9 to 9 T or -14 to 14 T, aligned along the [001] direction. 
In all cases, R$_H$ was positive, indicating hole-type carriers.
All electrical contacts were made using Epotek H20E conductive silver epoxy, with typical contact resistances between 1 and 3 $\Omega$.  

The structural phase transition was also observed by high resolution x-ray diffraction on beamline 2-1 at the Stanford Synchrotron Radiation Laboratory, using an incident x-ray energy of 13.0 keV. The diffracted beam resolution was set with a Ta-doped Si(111) crystal, which provides
a 2$\theta$ resolution of 0.01 degrees. Samples were coarsely crushed in a mortar and pestle and fixed to a zero-background quartz holder with GE varnish. Grinding the material thoroughly was found to induce an irreparable amount of strain in the lattice, evidenced by a reduced room-temperature lattice constant of 6.318 $\AA$ compared to that of coarsely crushed material (6.327 $\AA$) and suppression of the structural phase transition temperature below 50 K for Sn$_{.995}$Te, even when grinding was followed by a 6 hour anneal at 300 C.  Because the material was not finely ground, the intensity of each reflection varied significantly as a function of sample angle, $\theta$, for a fixed detector angle, 2$\theta$.  For each reflection, a value of $\theta$ resulting in high intensity was used. For the case of the indium-doped samples, a constant value of $\theta$ was used throughout the entire measurement.  However, during the measurement of the Sn$_{.995}$Te sample, this was not possible, because the structure was studied over a broader range of temperature. Thermal contraction of the sample holder forced different values of $\theta$ to be used at different temperatures to maintain reasonable peak intensity. For all samples, scans through $\theta$ -- $2\theta$ were taken, rocking $\theta$ by 2 degrees at each data point to average over a range of sample orientations. 

The lineshapes of the (420) and either the (640) or (620) peaks were measured as a function of temperature for each sample. The (400) peak was also measured as a reference of unsplit peak width as a function of temperature. Asymmetry was found in the lineshape of all peaks, attributed to strain in the sample. To account for this, a lineshape (640) or (620) measured above $T_{SPT}$ (95 K for $x = 0 \%$, 60 K for $x = 2.1 \%$, and 40 K for $x=3.4\%$) was fit to two peaks, each described by a linear combination of a Gaussian and a Lorentzian lineshape. Lower-temperature peaks were fit to two of these pre-defined asymmetric lineshapes, varying only the position and amplitude of each. Peak splittings were extracted as the difference between the maxima of the resulting two peaks.

Heat capacity was measured between 0.35 K and 5 K on 2-6 mg single crystals using the relaxation method with a Quantum Design Physical Properties Measurement System equipped with a helium 3 cryostat.  Single crystals were cleaved to provide a flat surface for good thermal contact to the sample platform.  Measurements were made in zero field and in an applied field of 1 T, to suppress the superconducting transition, at arbitrary orientations of the sample in the field. The electronic contribution, $\gamma$, was calculated from linear fits to $C / T$ vs T$^2$, for data taken in an applied field.

\section{Results}

\begin{figure}

\includegraphics[width=3.8in]{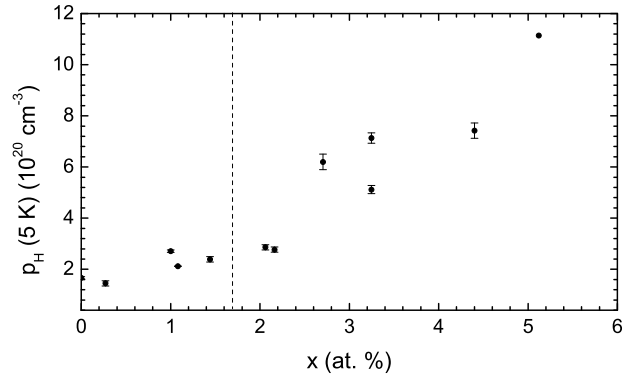}
\caption{\label{hall}Hall number, $p_H = 1/R_He$, at 5 K as a function of $x$ for Sn$_{1-\delta-x}$In$_x$Te.  The value of $p_H$ for $x=0$ indicates $\delta = 0.52(1) \%$.  The vertical dashed line indicates $x_c = 1.7(3) \%$.}
\end{figure}

Tin telluride is a multiband semiconductor,\cite{tung1969} but the Hall number, $p_H = 1/R_He$, shown in figure \ref{hall} as a function of indium content, $x$, can be used as a reasonable estimate of the carrier concentration.\cite{bushmarina1984} Uncertainty in the actual carrier concentration does not affect the following analysis. The value of $p_H$ for undoped SnTe is $1.66(4)\times10^{-20} cm^{-3}$, suggesting a value for $\delta$ of $0.52(1) \%$ (figure \ref{hall}).  As $x$ is well below the solubility limit of $\sim 20 \%$ and all materials were grown in similar growth conditions, it is likely that this value does not change significantly across the range of $x$ investigated here.  Additionally, EMPA measurements confirm that the Sn vacancy fraction is consistent throughout the series, to within 1 atomic percent. 

\begin{figure}
\centering
\includegraphics[width=3.5in]{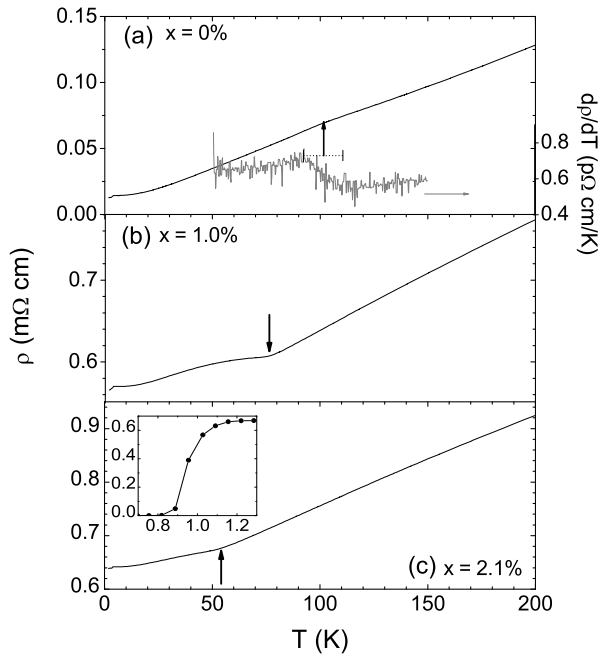}
\caption{\label{resistivity}Temperature-dependence of the resistivity from 1.8 K to 200 K for three representative indium concentrations.  Panel (a) also shows the derivative of the resistivity in the vicinity of the SPT.  The inset to panel (c) shows the superconducting transition for the sample with 2.1 \% indium.}
\end{figure}

\begin{figure}
\centering
\includegraphics[width=3.5in]{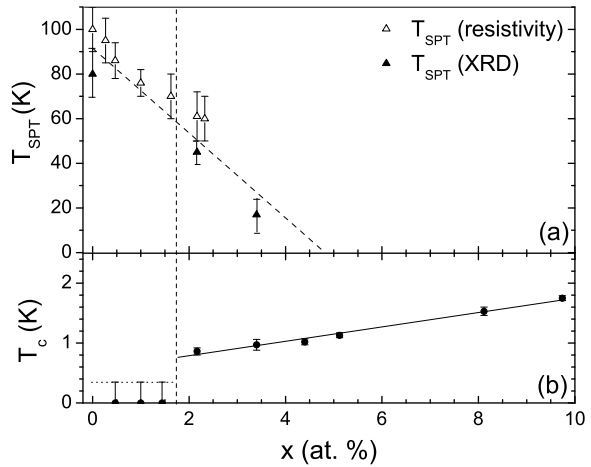}
\caption{\label{tc}(a) Critical temperature of the structural phase transition, $T_{SPT}$, with data points extracted from resistivity measurements shown in open triangles and data from high resolution x-ray diffraction measurements shown in filled triangles. The dashed line is a guide to the eye. (b) Superconducting critical temperature, $T_c$, determined from heat capacity measurements as described in the main text. The vertical dashed line indicates $x_c = 1.7(3) \%$.  The horizontal dotted line indicates the temperature limit of the instrument, and the solid line is a guide to the eye.}
\end{figure}

Resistivity data for samples with increasing In concentration are shown in figure \ref{resistivity} for three representative indium concentrations.  The derivative of the resistivity is shown for Sn$_{.995}$Te on the right axis of panel (a) to emphasize the anomaly at $T_{SPT}$.  For all In concentrations, the temperature of the structural phase transition was defined as the midpoint of the change in the derivative and is shown as a function of $x$ in figure \ref{tc}(a) (open triangles).  Error bars were determined from the width of the change in the derivative, as indicated in figure \ref{resistivity}(a).  The resistivity of undoped Sn$_{1-\delta}$Te shows a kink at this temperature (figure \ref{resistivity} (a)), agreeing with published data.\cite{kobayashi1976}  Indium doping both reduces $T_{SPT}$ and modifies the temperature dependence of the resistivity in the vicinity of the transition. For this reason, the resistive signal of the SPT could not be resolved for samples with In concentration, $x > 2.3 \%$.  Superconducting samples show a sharp drop in resistance at $T_c$, with a typical width of 0.1 K, from 10\% to 90\% of the full transition (inset to figure \ref{resistivity} (c)).  A small drop in resistivity at T $\sim$ 3.5 K appears in many samples.  This is attributed to superconductivity of Sn inclusions inherited from the melt growth and is suppressed in an applied magnetic field of $>$ 300 Oe, the critical field of metallic Sn. 

\begin{figure}
\centering
\includegraphics[width=3.5in]{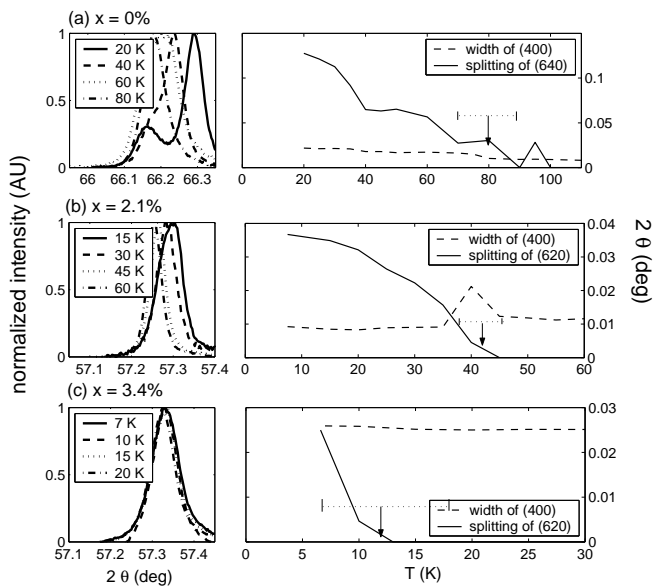}
\caption{\label{xrd} Left-hand panels show the temperature-dependence of the linewidth of the (a) (640) reflection for Sn$_{.995}$Te and of the (620) reflection for Sn$_{.995-x}$In$_x$Te with (b) $x$=2.1\% and (c) $x$=3.4\%.  Right hand panels show a summary of two-peak fits to these linewidths, as described in the main text.  Peak splitting of the (hk0) reflection is shown as a solid line, with the temperature-dependence of the width of the (400) peak, which is not affected by the structural transition, shown as a dashed line for reference.  Arrows indicate the temperature of the structural phase transition, with error bars shown as dotted lines, derived from this data as described in the main text.}
\end{figure}

Figure \ref{xrd}(a) shows the trend with temperature of the (640) reflection in Sn$_{.995}$Te.  The temperature dependence of the normalized peak lineshape is shown in the left-hand panel.  At 20 K, two peaks are clearly resolved.  Because the sample was only coarsely ground, preferential alignment is possible and no significance can be drawn from the relative peak intensities.  At 40 K, the smaller peak has become a shoulder; by 60 K, they have merged; and above 80 K, the full width at half maximum (FWHM) becomes constant at $2 \theta$ $=$ 0.06 deg, well above the experimental resolution of $2\theta = 0.01$ deg. The right hand panel shows the trend with temperature of the peak splitting of the (640) peak, derived as described above, along with the width of the (400) peak, for comparison. The (400) peak shows no significant change in linewidth in this temperature range. Figure \ref{xrd}(b) displays similar data for the (620) reflection  for $x=2.1\%$.  In this case, two peaks cannot be resolved at the lowest temperatures measured.  However, a clear trend in the FWHM is observable, and lower temperature peaks are fit better by two separate peaks than one single peak. For this sample, the FWHM becomes roughly constant at 0.04 deg $2\theta$ above 40 K. For $x=3.4\%$, shown in panel (c), the trend of the FWHM with temperature is less obvious, but an increase in the FWHM above $2\theta = 0.07$ deg is observable below 15 K, and the lowest temperature data can be fit well with two separate peaks. Material of indium concentration $x = 4.4 \%$ was also measured, and showed no variation in FWHM down to the base temperature of 7 K.

The SPT temperature shown in figure \ref{tc}(a) (solid triangles) was approximated as the temperature above which the two peaks had fully merged. Error bars were chosen to include the region in which the peak separation was less than the linewidth of the (400) peak.  Phase transition temperatures determined from x-ray diffraction measurements appear approximately 10 K lower in temperature than the transition extracted from resistivity measurements.  This difference is attributed to a combination of poor thermal contact with the sample through the quartz sample holder and uncertainty in determining the peak separation when the splitting was less than the FWHM, both of which would lead to a reduction in the observed value of $T_{SPT}$.

\begin{figure}
\centering
\includegraphics[width=3.5in]{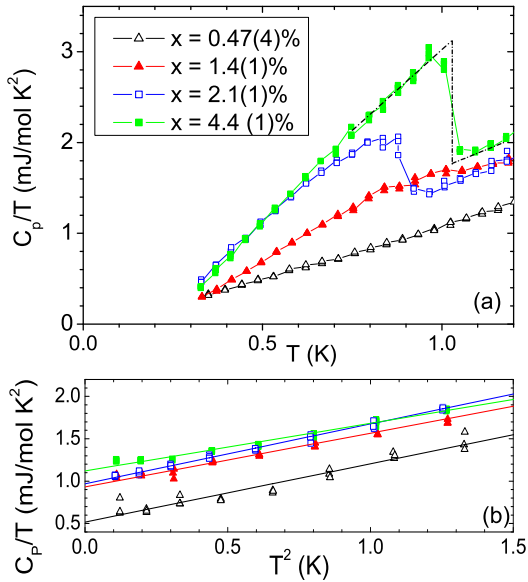}
\caption{\label{cp}(color online) (a) Heat capacity of Sn$_{1-\delta-x}$In$_x$Te at zero field in the vicinity of the superconducting phase transition for four representative indium concentrations. Dashed line shows geometric construction used to extract $\delta C$. (b) C/T vs T$^2$ for the samples in an applied field of 1 T, showing linear behavior.  Lines are linear fits to data from $T^2 = $ 0 to 1.5 K$^2$.}
\end{figure}

Heat capacity data showing the superconducting phase transition are shown in figure \ref{cp}(a) for representative samples. For samples of doping 4.4(1)\% (solid squares) and higher, a sharp transition is observed at $T_c$.  For $x=2.1(1)\%$ (open squares), some broadening in the transition is observed. For $x=1.4(1)\%$ (filled triangles), some remnant superconductivity is left, but no bulk superconductivity is observed. For $x=$ 0.47(4)\% (open triangles), no trace of superconductivity is observable.  This gradual broadening of the superconducting transition as $x$ approaches $x = $$x_c$ is indicative of a sharp threshold from nonsuperconducting to superconducting behavior near this indium concentration. While the degree of inhomogeneity is likely to be consistent throughout the series, a small change in $x$ across the sample will lead to a greater variation in $T_c$ for $x$ near $x_c$ than for $x$ far from $x_c$. 

The superconducting transition temperature was extracted as the midpoint of the discontinuity in specific heat and is shown as a function of $x$ in figure \ref{tc}(b). Error bars were defined as 10\% to 90\% of the full transition height.  Values of $T_c$ extracted in this way were approximately 0.5 K lower than the midpoint of resistive transitions. Since heat capacity measurements are inherently a more accurate measurement of bulk behavior than a transport measurement, we use values of $T_c$ extracted from heat capacity measurements to define $T_c$. For indium concentrations less than 1.4(1)\%, no superconductivity is observed down to 0.35 K, consistent with the behavior of indium-free Sn$_{0.995}$Te reported elsewhere.\cite{allen1969} Above $x =2.1(1)\%$, a sudden increase in $T_c$ is observed.  Taking an average of these two concentrations, we estimate a critical In concentration for superconducting behavior, $x_c = 1.7(3)\%$.  For $x >$ $x_c$, $T_c$ rises approximately linearly with $x$ from 0.8 K to 1.8 K for $x = 9.9(1) \%$.  These $T_c$ values are an order of magnitude higher than those found in the pure Sn$_{1-\delta}$Te system, and are consistent with previously published data for polycrystalline Sn$_{1-\delta-x}$In$_x$Te.\cite{bushmarina1986}

\begin{figure}
\centering
\includegraphics[width=3.5in]{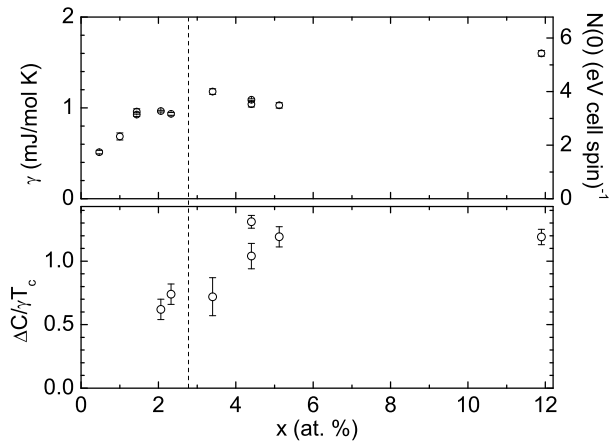}
\caption{\label{gamma}(a) Electronic contribution to the specific heat, $\gamma$ (left axis), with corresponding values for the density of states per spin, $N(0)$ (left axis). (b) Magnitude of the superconducting anomaly in the heat capacity, $\Delta C/\gamma T_c$.  The upper axis is the value of the BCS prediction, $\Delta C/\gamma T_c = 1.43$.  The vertical dashed line indicates $x_c = 1.7(3) \%$.}
\end{figure}

The electronic contribution to the specific heat, $\gamma$, was obtained from a linear fit to C/T vs T$^2$ at a field of 1 Tesla, for which $T_c$ is suppressed below 0.3 K, shown in figure \ref{cp}(b). Error bars were determined from the resolution of the mass measurement and uncertainty in the linear fit. The size of the anomaly, $\Delta C$, was estimated by extrapolation of the data from above and below $T_c$, a typical example of which is shown as a dashed line in the upper panel of figure \ref{cp}. Error bars were estimated by comparing linear extrapolations to C/T over different temperature ranges above and below $T_c$. Figure \ref{gamma} shows (a) $\gamma$ and (b) $\Delta C/\gamma T_c$ as a function of $x$.  The magnitude of $\Delta C/\gamma T_c$ is near the BCS weak coupling value of 1.43 for high indium concentrations, suggesting that Sn$_{1-\delta-x}$In$_x$Te is a weakly coupled superconductor.  A trend away from this value at low concentration is observed, similar to that previously reported in Pb$_{1-x}$Tl$_x$Te.\cite{yana}  This behavior may reflect inhomogeneity in the In concentration.

\section{Discussion}

Given that the value of $\Delta C/\gamma T_c$ was found to be consistent with that of a weakly-coupled superconductor, it is reasonable to work in this limit. In the BCS weak coupling limit, the variation of $T_c$ with the density of states at the Fermi level, $N(0)$, is given by the equation
\begin{equation}
kT_c=1.13\hbar \omega_c e^{-1/N(0)V}
\end{equation}
where $V$ is the strength of the average electron-electron pairing interaction, and $\omega_c$ is a cutoff frequency, which, for a conventional phonon-mediated superconductor, is usually given by the Debye frequency.\cite{tinkham}  

\begin{figure}
\centering
\includegraphics[width=3.5in]{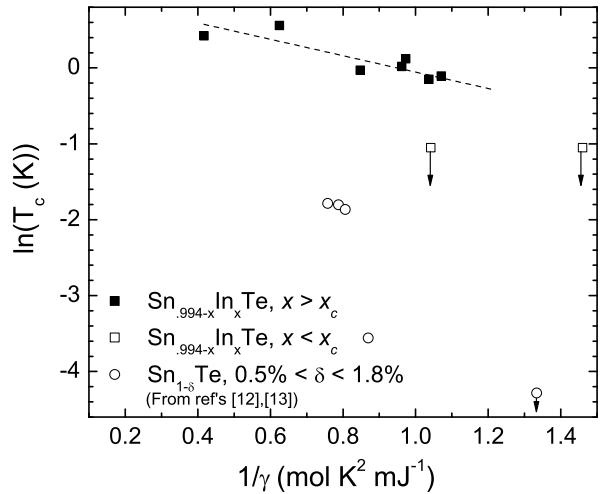}
\caption{\label{lntc}Comparison of the inverse of the electronic contribution to the specific heat, $1/\gamma$, proportional to $1/N(0)$, to the natural log of $T_c$ for indium-free Sn$_{1-\delta}$Te (open circles) (from ref. \onlinecite{hulm1968,phillips1971}), and indium-doped tin telluride with $x$ $>$ x$_c = 1.7(3)\%$ (closed squares) and $x$ $<$ $x_c$ (open squares).  Dashed lines depict linear fits to the appropriate group of data. Arrows below data points indicate a lack of superconductivity down to the instrumental limit.}
\end{figure}

Assuming a constant value of $V$ with $x$, a plot of $ln(T_c)$ vs 1/$\gamma$ will yield a straight line, where the slope is related to$1/V$, and the intercept is related to $\omega_c$. The results of this analysis for Sn$_{.995-x}$In$_x$Te samples from this work are shown in figure \ref{lntc}, with samples with $x$ $>$ $x_c$ = 1.7(3)\% shown in filled squares.
 The data with $x >$ $x_c$, can, indeed, be fit by a line with slope of $\sim$-1.0(2) mol K$^2$ mJ$^{-1}$, suggesting a BCS pairing interaction strength of $V$ $\sim 0.3$ eV$\times$(unit cell)$\times$ spin.  The dimensionless parameter $N(0) \times V$ corresponds to the magnitude of the electron-electron interaction, and can be decomposed as $N(0) V = \lambda - \mu^*$, where $\lambda$ is the attractive electron-phonon coupling strength, and $\mu^*$ represents the screened coulomb repulsion.\cite{poole}  In this case, this extracted estimate of $V$ implies that $\lambda - \mu^* =$(0.3 eV$\times$ cell $\times$ spin)$ \times $(3(eV cell spin)$^{-1}) = 0.9$.  Typical values of $\mu^*$ for other superconductors are between 0.1 and 0.2,\cite{poole} suggesting that $1 < \lambda < 1.1$, similar to superconductors of intermediate to strong coupling strength, such as niobium or lead, which both show a $\Delta C/\gamma T_c$ of order 2 or larger.\cite{poole}  This is in direct contradiction to the weak coupling value of $\Delta C/\gamma T_c < 1.43$ found in the present work.  Additionally, the intercept of this plot can yield the value of the cutoff frequency used equation (1).  Such analysis reveals an unphsically small value of $\omega_c = 2.4 K$. These discrepancies show that $T_c(x)$ cannot be accounted for solely by variation in the density of states, for instance, due to formation of a narrow impurity band. A similar result has been found previously, in the case of Pb$_{1-x}$Tl$_x$Te.\cite{moizhes1991} Rather, it seems that, for $x > x_c$, indium impurities in tin telluride must introduce a variation of $V$, and possibly of $\omega_c$, with $x$.


Values of $T_c$ for $x < x_c$ were undetermined, but below 0.35 K.  This upper limit is indicated in figure \ref{lntc} by open squares. Data for undoped Sn$_{1-\delta}$Te collected from references \onlinecite{hulm1968,phillips1971} are shown in open circles. While there are not enough well-determined data points to fit a line to the data for the material with $x < x_c$, these data clearly show a weaker coupling constant than that of the material with higher indium content. This indicates that, within this model, indium-doped samples with $x$ $>$ $x_c$ show a pairing interaction strength much greater than that of the indium-free Sn$_{1-\delta}$Te, and of Sn$_{.995-x}$In$_x$Te with $x$ $<$ $x_c$, even when the two materials have a similar $N(0)$.

Given the presence of the ferroelectric SPT in the host material,\cite{kobayashi1976,katayama1980} softened phonons related to this structural transition are, at first sight, a likely candidate for the source of this enhanced pairing. In conventional phonon mediated superconductors such as 
V$_3$Si,\cite{testardi1974} it has been shown that the presence of a structural phase transition in close vicinity to the superconducting critical temperature enhances superconducting pairing through softening of phonon modes involved in Cooper pairing.  This effect has been investigated in systems of SnTe:GeTe:PbTe thin film alloys, where a link between the ferroelectric Curie temperature and the superconducting critical temperature was found.\cite{grassie1972, benyon1973} However, in this work we find that, while $T_c$ does increase as $T_{SPT}$ decreases, the sudden jump in $T_c$ at $x_c = 1.7(3) \%$ is not correlated with a comparable decrease in $T_{SPT}$.  Furthermore, high carrier density samples of undoped Sn$_{1-\delta}$Te with similarly low values of $T_{SPT}$ do not superconduct until well below temperatures of 0.1 K.\cite{kobayashi1976,allen1969}  Even when the structural phase transition is fully suppressed in Sn$_{1-\delta}$Te, superconductivity remains below 1 K until the carrier concentration is increased a full two orders of magnitude above the carrier concentration of these indium-doped materials.\cite{kobayashi1976,allen1969}  This lack of correlation between $T_{SPT}$ and $T_c$ suggests that soft phonon modes associated with the ferroelectric phase transition cannot explain the anomalously high $T_c$ value in highly doped samples of Sn$_{0.995-x}$In$_x$Te.

Rather, we suggest that unique electronic properties of the indium dopant atoms are responsible for this pairing enhancement. Previously published measurements of the Hall carrier density as a function of Sn vacancy density, $\delta$, demonstrate that, when $\delta$ is less than or equal to approximately half the indium concentration, $x/2$, the carrier concentration becomes insensitive to changes in $\delta$.\cite{bushmarina1984} The authors of reference \onlinecite{bushmarina1984} interpret this as a pinning of the Fermi level in a narrow band created by the indium impurity states. For $\delta = 0$ and $x > 1\%$, the indium impurity states are half filled, with an average valence state of $+2$. Additional tin vacancy doping introduces two holes per vacancy, emptying electrons from the localized indium impurity atoms. Thus, at a value of $\delta$ $\sim$ $x/2$, the indium impurity sites become completely emptied of electrons, and for larger values of $\delta$ the Hall concentration begins to change again. However, when the concentration of indium impurities, $x$, is varied, with fixed $\delta$, the Hall concentration is observed to change (figure \ref{hall}), even in the region where the Fermi level is pinned.  This is attributed to a change in the energy of the indium impurity states as a function of $x$.\cite{bushmarina1984}

In the present work, the indium concentration above which enhanced superconductivity is observed, $x_c$ = $1.7(3) \%$, is consistent with that demonstrated in reference \onlinecite{bushmarina1984}, where pinning of the Fermi level is found for $\delta \sim 0.5 \%$, the Sn vacancy density of the samples reported here. Comparison of the data shown in figure \ref{hall} to the data published in reference \onlinecite{bushmarina1984} suggests a critical indium concentration for Fermi level pinning in the range of $ 1 -– 2.5 \%$, where the present work finds the enhancement of $T_c$.  This observation is in agreement with previously published measurements that suggest that enhancement of superconductivity only occurs when the Fermi level is pinned.\cite{bushmarina1986}  It has been suggested that the $T_c$ enhancement found in the region of Fermi level pinning is a result of the enhancement of the density of states, $N(0)$, due to the formation of a narrow impurity band.\cite{bushmarina1986}  In the present work, however, we find that the relationship between $T_c$ and $N(0)$ is not the same for samples with In concentration, $x$, above and below $x_c$, suggesting that an enhancement of $N(0)$ in the narrow impurity band is not sufficient to explain the order of magnitude increase in $T_c$ when $x > x_c$.  Instead, it appears that the pairing interaction itself is fundamentally stronger for In-doped SnTe, when $x$ exceeds $x_c = 1.7(3)\%$.

A similar link between superconductivity and Fermi level pinning has recently been demonstrated for the closely related material Pb$_{1-x}$Tl$_x$Te.\cite{yana, joerg}
 Correlated with both Fermi level pinning and superconductivity in Pb$_{1-x}$Tl$_x$Te is the presence of a low temperature upturn in resisitivity, similar to a Kondo effect, though magnetic impurities were ruled out as a source.\cite{yana}  This was attributed to a charge-Kondo effect.\cite{joerg} In this model, fluctuations between $+1$ and $+3$ Tl valence states, made degenerate by the pinning of the Fermi level, lead to a Kondo-like resistive upturn at low temperatures, much as fluctuations between degenerate up and down spin states lead to enhanced low temperature scattering for magnetic impurities in a metal in a conventional Kondo model.  These local charge fluctuations have been shown theoretically to provide an electronic source of Cooper pairing in models of negative U superconductivity.\cite{moizhes1981,shelankov1987} Electrons tunneling onto and off of impurity sites are encouraged to do so in pairs by the high energy of the intermediate $+2$ valence state, compared to the filled-shell $+1$ and $+3$ states.  In contrast to the model used to explain the data in reference \onlinecite{bushmarina1984}, this model relies on a localization of electrons on the thallium impurity site, in the absence of an impurity band.  Recent ARPES measurements showing a lack of a narrow band at the Fermi level in Pb$_{1-x}$Tl$_x$Te for $x > x_c$ appear to support this model.\cite{nakayama}

In the case of Sn$_{1-\delta-x}$In$_{x}$Te, a mixed-valence scenario for Fermi level pinning similar to that proposed for Pb$_{1-x}$Tl$_x$Te might better explain the data shown in figure \ref{lntc}. In this model, the enhancement in pairing strength observed for $x > x_c$ would arise from negative U pairing on the indium impurity sites, and hence vary with the number of impurities. While no direct probe of the indium valence in Sn$_{1-\delta-x}$In$_x$Te has been investigated to date, the tendency for indium to skip the $+2$ valence makes this material an excellent candidate for charge-Kondo behavior.  Such a mixed-valent state has been demonstrated in Pb$_{1-x}$In$_x$Te using x-ray photoelectron spectroscopy.\cite{drabkin1982} While a mixed valent state is likely by analogy in Sn$_{1-\delta-x}$In$_x$Te, the effect of the structural phase transition on the low temperature resistivity obscures the search for any subtle Kondo-like effects in these samples.  For this type of study, samples of much higher Sn vacancy density must be used, in order to suppress the SPT completely before doping indium into the material.

\section{Conclusions}
Single crystal studies of the superconducting degenerate semiconductor Sn$_{1-\delta-x}$In$_x$Te have been presented. Analysis of the variation of $T_c$ with the density of states, $N(0)$, reveals two regions of superconducting behavior. For $x$ $<$ 1.7(3)\%, a superconducting pairing strength not inconsistent with that of the parent compound, Sn$_{1-\delta}$Te, is observed, while samples of higher indium content show an enhancement of $T_c$ relative to indium-free samples of similar density of states.  This rules out models relying on enhancement of the density of states at the Fermi level due to formation of a narrow impurity band to explain the enhancement of $T_c$ on adding indium.  In addition, measurement of the effect of indium on the temperature of the ferroelectric structural phase transition, $T_{SPT}$, rule out softened phonons as a cause of this $T_c$ enhancement.  These results establish Sn$_{1-\delta-x}$In$_x$Te as a candidate negative U superconductor.

\section{Acknowledgements}
The authors would like to thank Robert E. Jones for technical assistance with EMPA measurements, Doug Scalapino for his assistance with theoretical discussions, and  Mikhail Kerzhner for assistance with crystal growth and characterization.  This work is supported by the DOE, Office of Basic Energy Sciences, under contract no. DE-AC02-76SF00515.  Portions of this research were carried out at the Stanford Synchrotron Radiation Laboratory, a national user facility operated by Stanford University on behalf of the U.S. Department of Energy, Office of Basic Energy Sciences.

\end{document}